# Flares as fingerprints of inner solar darkness

Konstantin Zioutas [1,2], Mary Tsagri [1,2],
Yannis Semertzidis [1,3], Thomas Papaevangelou [4]

1) University of Patras, Patras, Greece
2) CERN, Geneva, Switzerland
3) Brookhaven National Laboratory, USA
4) DAPNIA, CEA-SACLAY, Gif-sur-Yvette, France

Emails: zioutas@physics.upatras.gr ; mary.tsagri@cern.ch ;
yannis@bnl.gov ; thomas.papaevangelou@cern.ch

**ABSTRACT**

X-ray flares, whose primary process we still miss, and other much weaker solar brightenings have their roots in magnetized regions. Until now, solar X-ray emission had actually been discarded as a potential axion signature, as it did not match two basic expectations of the solar axion model: a) an axion X-ray signal must appear exclusively near the disk centre, and b) its analog spectrum must peak at ~4-5 keV. On the contrary, we show here that they can appear in all longitudes and their intensity can peak at low energies. Due to Compton scattering off the (plasma) electrons, the outward propagation of X-rays from axions or the like converted near the Sun's surface can explain the observed low X-ray energy distribution and its non-directivity. Simulation points at the photosphere as the birth place of the presumed axion conversion, implying an axion rest mass $m_a \approx m_\gamma \approx (1-2) \cdot 10^{-2} eV/c^2$. This result supports previous claims, whose strength scales with $B^2$, typical for axions. At present, even optimistic parameter values cannot reproduce the measured X-ray intensities, i.e. either the solar axion source is not as anticipated or/and the inverse Primakoff-effect is not the main interaction mode. As a generic example we mention axion conversion in magnetic field gradients. The simulated photon spectrum peaking at low energies matches (qualitatively) reconstructed active and quiet Sun spectra. Interestingly, the shape of a reconstructed solar photon spectrum for the flaring Sun fits an energetically squeezed X-ray spectrum from converted axions. The recently observed hard solar X-ray emission from the quiet Sun could be alternatively due to massive and/or light axion involvement.

## 1. Introduction

Solar axion helioscopes à la Sikivie [1] (e.g. the running CAST and SUMICO [2]) utilize strong macroscopic magnetic fields in order to force radiatively decaying particles, like the celebrated axions, to coherently transform to photons that can be detected downstream with high efficiency and accuracy. Since solar X-ray flares (from the largest ones to the weakest brightenings) and other X-ray emission happen to be associated preferentially with magnetic places, and, taking into account that their trigger mechanism has been unknown for 150 years ("...we still miss the primary process ..." [3]), they should have been recognized long ago as the first candidate-signals for solar axions. However, they were discarded from further consideration, since the measured properties did not match basic requirements of the standard solar axion model: a) their analog X-ray spectrum does not peak at ~4-5 keV. Instead, it is increasing towards lower and lower energies, and more importantly, b) the flaring Sun appears equally in all longitudes between the west and the east solar limb, while X-rays from converted axions should emerge radially outwards, distinguishing thus the solar disk centre. Reference [4] gives an example how pseudoscalars in general are expected to behave when strong magnetized sunspots cross the disk centre, applying the working principle of an axion helioscope.

In this work we show that both 'prejudices' against axion involvement in the X-ray bright Sun can be overcome, opening a new window of opportunity while studying our nearest star. From the axion point of view, the question has been investigated as to whether the huge solar surface magnetic fields can be a much better axion-photon catalyst due to some occasionally built-in fine-tuned enhancement, whatever the actual mechanism. This could explain certain mysterious solar observations (as argued in refs. [5-7]), but also otherwise overlooked solar X-ray emission as coming from converted axions, e.g. via the coherent inverse Primakoff-effect, with the intensity following an axion-characteristic $B^2$-dependence. Other known parameters in axion physics can also reinforce the axion conversion. For example, the plasma frequency of a magnetized volume [8] in the restless Sun may be only occasionally "tuned" to the axion rest mass. This could of course favour temporal and local axion conversion, resulting eventually in a plethora of observed transient or almost steady X-ray emission. Furthermore, previous work motivated by the very steep transition region (TR) separating the chromosphere and the corona [9-11] addressed the steady solar X-ray emission as coming from gravitationally trapped massive axions of the Kaluza-Klein type, whose spontaneous decay near the Sun gives rise to a self-irradiation of the whole solar atmosphere, implying an inwardly directed radiation pressure. This can explain the otherwise serious and nagging problems with the TR [12] and eventually also the measured elemental abundance anomalies from the quiet Sun (see below).

As happened in the past with new (solar) physics, it is not at all obvious that we now know *all* the naturally occurring enhancement mechanisms that might take place in the nearby Sun, and which could materialize much more efficiently the celebrated axions or other as yet (un)predicted exotica to photons [13]. It is worth clarifying here that if one takes the experimentally derived limits for the axion-to-photon coupling strength (see e.g. ref. [2]), a solar X-ray luminosity from QCD-inspired axions should be at least some 20 orders of magnitude below that of the luminous Sun (=$3.8 \cdot 10^{33}$erg/s), and this can not be detected at present. Therefore, an enhancement by a factor of ~$10^9$ must be somehow at work in the Sun, or/and, an unforeseen interaction strength has been overlooked before. In both cases, there is room for surprises. As a very recent example, we mention a novel concept suggested by Guendelman [14] about axion interaction with magnetic field *gradients*, in spite of the fact that the interaction strength has not been quantified as yet. This magnetic field configuration was never taken into account in an axion experiment, at least not on purpose. Surprisingly, and to the best of our knowledge, solar X-ray activity is in fact closely related also to places with strong magnetic field gradients [15], and therefore this suggestion is even more of potential interest for the justification of this work, since it cannot reproduce quantitatively the intensity. After all, this new concept may further contribute to the dominant role magnetic fields (can) play as axion-photon catalysts. Thus we seem to live in a challenging situation, since both the actual interaction mode and the nature of the solar exotica in question may be different from what has been widely anticipated so far.

For reasons of simplicity, we refer mainly to the solar axion scenario, which stands throughout this work also for any other particle candidate with similar properties [13]. Thus, the (elastic) coherence axion-to-photon oscillation implies that the emerging photon is collinear with the incoming axion. This is a cornerstone in axion-telescopy, and it is just this property which allows the operation of axion telescopes of the Sikivie type with extremely high space resolution. Owing to the implied axion-photon collinearity, for an outside observer the detected X-ray properties from such a process near the Sun's surface can be used to trace back very precisely their place of birth (e.g., with an orbiting X-ray telescope with high angular resolution). Since the solar axion source (=solar core), the intervening surface transverse magnetic field component, and the X-ray observer define a straight line, the solar disk centre is clearly distinguished within the axion framework, which is actually contrary to everyday experience with solar X-ray data. This has far-reaching implications for axion identification in solar X-rays. According to this scenario, X-rays from magnetically converted solar axions, i.e., coherent inverse Primakoff-effect, can be observed only from a spot near the disk centre as small as the hot solar core (~10-20% of the solar radius) [4]. This reasoning alone was actually sufficient to exclude as self evident the standard solar axion scenario from being behind X-ray emission in the rest of the magnetic Sun surface, where the bulk of the X-ray activity takes place.

For example, all but two [15,16] of the previous claims were located outside the disk centre, and therefore they also should have been rejected from further consideration following the reasoning of ref. [4], which addressed only axion-like particles, i.e., ALPs in present day jargon. Referring to axions, ref. [4] is cited nevertheless quite often in ongoing work with data from Yohkoh, RHESSI, Hinode, etc. The 'prejudice' was and is thus that magnetically converted solar axions could show up only at the disk centre region. The actual motivation of this work was to find a way to surpass this otherwise very serious constraint for most of the Sun's X-ray bright surface [5-7]. Still, the reasoning of Ref. [4] can be valid, but conditionally, as it applies to low rest-mass pseudoscalars with relatively large coupling constant, which convert high in the upper chromospheres or beyond. It is certainly not applicable in a wide parameter phase space of the dynamic solar atmosphere and the standard axion.

Therefore, the main purpose of this work is to show that magnetically converted axions can be visible even from the whole solar disk for an Earth X-ray observer, and, the measured analog spectrum can be different from the original axion spectrum, as it can be shifted towards lower energies. Interestingly, this is in accordance with observation. We simulate below the propagation of converted axions to X-rays near the photosphere. The obtained erase of directivity and the energy degradation are both independent on the actual axion interaction mechanism involved, which must be, however, enhanced compared to what we know about QCD-inspired solar axions. The reasoning of this work supports also previous specific axion claims [5-7], which were based mainly on the $\sim B^2$-dependence of X-ray emission, or indirectly, on the measured increase of elemental abundance above magnetic pores [=small sunspots], which show also a striking $B^2$-behaviour.

## 2. Isotropic X-ray emission from converted solar axions

In the following we focus on a hot X-ray flare region. However, the same scenario (scaled down somehow) might also apply to a much weaker solar X-ray brightening / emission. The trigger of all these events is actually often unknown [15,17], and therefore solar activity is unpredictable and often mysterious. Whatever the configuration of the magnetic fields, it is known from observations that they are spatiotemporally associated with flaring activity; they can work, at least partly, not as the usually assumed, but still unspecified, magnetic energy reservoir for solar flare activity, but as the catalyst for axion-to-photon conversion. This is in fact the main difference between the pure magnetic and pure axion related explanation for the same end result. Thus, conventional and new physics can coexist and be complementary without necessarily excluding each other.

Appropriate environmental parameter values naturally allow for an axion interaction to take place much more efficiently. Even though it is beyond the scope of the present work to explain specific solar observations, we mention here in short that, within the suggested axion inspired scenario, in a flare:

a) photoionization can take place with converted axion-like exotica near the magnetized solar surface, which can irradiate the overlying layers to a plasma (see footnote 1). In the quiet Sun, it is only the tiny layer below the transition region to the deep photosphere, which is actually not ionized, distinguishing this layer from the rest of the Sun.

b) the created ~4 keV photo-electrons can solve the 'electron number problem' in flare models [15,18], while their bremsstrahlung intensity is negligible..

c) the associated radiation pressure due to X-rays from converted axions could also be behind the closely related Coronal Mass Ejections (CMEs), e.g., via a collective type ion acceleration concept of the electron cloud (see section 2.1.1 (1) in ref. [9]). Note that, the CME protons have a velocity of ~$10^{-3}$c, which corresponds to the one of an electron cloud in the sub-eV range, while ~4 keV electrons start initially with a speed of ~0.1c. Such velocities leave room to account for inefficiencies due to the not ideal conditions for a collective acceleration.

Furthermore, the X-ray flare surface brightness does not actually exceed that of the quiet Sun luminosity even for the strongest events, while we know observationally that the flare region is heated up to 10-30 MK [15,18]. Whatever the reason behind this high temperature, it is remarkable that it coincides with that of the core ~700 000 km underneath. Therefore, such a hot region remains anyhow as a fully ionized plasma at least until it starts cooling down to ambient pre-flare temperatures. X-rays from converted exotica [be it via the inverse Primakoff-effect inside the magnetic field, or the axion interaction with the magnetic field gradient, or any other as yet unforeseen mechanism(s)] do interact with the surrounding plasma electrons (Compton effect). The scattering probability [19] is about 50% for an equivalent (ionized) hydrogen column density of 1-2 g/cm$^2$.

Interestingly, such column densities exist near the solar surface. For example, they are: ~4.4 g/cm$^2$ at the surface of the photosphere (increasing rapidly underneath), ~1 g/cm$^2$ and ~$10^{-3}$g/cm$^2$ at +200 km and +1000 km into the chromosphere, respectively, while the plasma density in the solar corona changes dynamically by a factor of 10-100 at any given time [20]. Then, occasionally, the isotropic Compton scattering of X-rays from converted axions at the upper atmospheric plasma can still be quite considerable, i.e., also at larger heights than one should assume for the static atmosphere. Thus, if the actual flare trigger place is (far) below the Transition Region (at ~2000 km), the initially radially and outwardly emitted X-rays from out streaming and converted axions can also photoionize the intervening neutral gas above [1].

Note that it is just this plasma above the flare trigger place, whatever its origin, which can act as the isotropic Compton X-ray scatterer (see below). Then, in such a case, we have to deal with a kind of a dynamic "solar surface effect", whose thickness can be as much as (few) 1000 km near the solar surface. The next section gives the results of a first Monte Carlo calculation.

---

[1] If only converted axions are the source of ionization, this might last for some time. The estimated time to photoionize ~2 g/cm$^2$ above the flare trigger place is of the order of $10^4$ s, assuming an axion originated solar X-ray surface brightness of ~$10^{-3}$ L$_{solar}$ to be the cause. This requirement is in principle possible, even though extreme, but a large solar flare is also an extreme event. Or, some other conventional reaction mechanism can be in synergy, e.g. like the celebrated reconnection of opposite magnetic fields, which imply also a magnetic field gradient along the neutral line, where solar X-ray activity is associated with.

## 3. Simulation

We show here the results of a Monte Carlo simulation for the propagation of X-rays from converted QCD-type solar axions at the solar surface magnetic field. In this simulation with the CERN Geant4 code, the photoelectric effect was inactivated, in order to mimic the propagation of X-rays in a thick plasma (> few $g/cm^2$), which finally escape into free space by a "random walk". The striking isotropic X-ray re-emission derived from this simulation for the two column densities given in Figure 1, supports the basic idea behind this work. This means that the memory of the initial axion trajectory, taking entirely by the first photon coming out of the inverse Primakoff effect, is statistically erased at the level of 90% (4.4 $g/cm^2$) and 99.9% (16 $g/cm^2$).

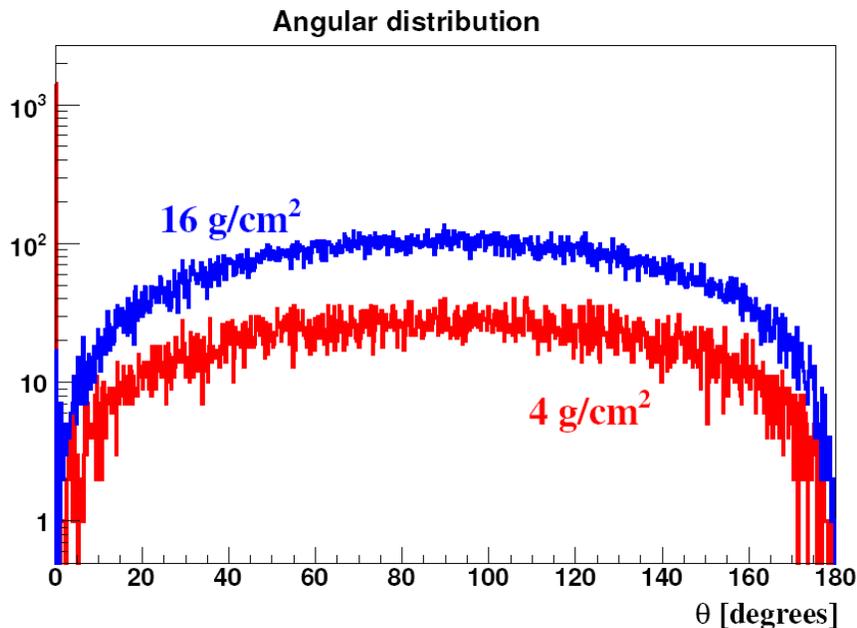

**FIGURE 1**   Simulation with the CERN/Geant4 code of the angular distribution of converted solar axions X-rays (dotted line in Fig. 2) inside the magnetized surface for two different column densities of the solar atmosphere above the initial place of birth of the X-rays. The photoelectric effect has been inactivated resembling thus free plasma electrons. The direction of the radially out streaming axions is at $\Theta=0°$. Simulated converted axion events = 16415. Number of escaping (=not interacting) X-rays at $\Theta = 0°$: 1422 or 8.7% (4 $g/cm^2$), and, 18 or 1.1‰ (16 $g/cm^2$).

With the same Monte Carlo calculation, we have also followed the photon energy loss after each Compton scattering. The surprising results are shown in Figure 2. The original X-ray spectrum assuming conventional QCD-inspired axions, which have been converted at the solar surface, is shown as a dotted line (**a**). The resulting analog spectra after the X-ray 'random walk' through 4g/cm$^2$ (**b**), 16 g/cm$^2$ (**c**) and 49 g/cm$^2$ (**d**) demonstrate the onset of energy degradation with increasing column density, due to apparently increasing multiple inelastic Compton scatterings. Note that the basic Geant4 code we used for this simulation has a photon threshold at 1 keV. Apparently, the actual spectra peak stronger towards even lower energies than shown in the simulated ones of Figure 2, and this might also apply to much more faint solar X-ray emission including the quiet Sun (see Figure 2). Interestingly, this Monte Carlo calculation further shows that bremsstrahlung from some keV electrons becomes redundant as the low energy photon source. The requirement for such a hot electron cloud resulted to the "electron number problem" of flares. Due to the small yield of the process (~$10^{-5}$ per electron), the required electron flux becomes energetically a serious problem for modelling [22]. This is no longer the case for the small energy loss of hard X-rays appearing during a flare, introduced through multiple Compton scattering (in our suggestion converted axions are the actual X-ray source).

**Numerical example for axion-to-photon conversion**

If the plasma density matches the axion rest mass, i.e., $m_\gamma=\hbar\omega_{plasma}\approx m_{axion}c^2$, then the axion-to-photon oscillation length [=coherence length] becomes quasi infinite. In reality, there is always a density inhomogeneity as happens strongly with the Sun [23], which constraints this coherence length. With the pioneering work of Van Bibber and coworkers [8] and the actual equation

$$\frac{\Delta\rho}{\rho} = 2\frac{\Delta m}{m_\gamma}$$

the relevant relation for $\rho \neq$ constant is:

$$\frac{\Delta\rho}{\rho} = \frac{4\pi E}{L\,m^2}$$

(m=m$_{axion}$). Assuming the static solar surface density profile [23] with $\Delta\rho/\rho \approx 10^{-2}$, we arrive at a coherence length of L ≈3 km and E ≈ 4 keV for the solar surface with $\rho\approx 2\cdot 10^{-7}$ g/cm$^3$ (i.e., for $\hbar\omega_{plasma}$= m$_{axion}$c$^2$ ≈ $10^{-2}$ eV/c$^2$). In the following we make a rather optimistic but still realistic estimate. Thus, we can have a rough idea about the order of magnitude of the effect one may expect (optimally?) near the solar surface, where the axion-to-photon conversion must occur, for the scenario of this work to apply. Even if local solar dynamical behaviour allows to keep the spatial relative density change within $\Delta\rho/\rho \approx$ 1% over a distance which roughly corresponds to 10 times that of the static solar density change ($\Delta\rho/\rho \approx$ 10 %), the corresponding coherence length becomes L≈20 km. Furthermore, assuming a local field of 1 Tesla (note 0.5 Tesla have been measured already at some sunspots) and a coupling constant $g_{a\gamma\gamma}\approx 10^{-10}$ GeV$^{-1}$, the estimated axion conversion efficiency [2] is still very small: $P_{a\to\gamma}\approx 10^{-12}$. For comparison, a very large luminous flare requires instead an efficiency of the order of $10^{-3}$, implying a missing factor of ~$10^9$. For less extreme events like microflares, nanoflares, flaring brightenings, or even non-flaring X-ray emission from active regions, the missing factor would be much smaller. Following this reasoning, even a further improvement by a few orders of magnitude is still insufficient. This suggests to either consider the existence of other axion-like exotica, or to speculate that much stronger fine tuning by the restless Sun is at work.

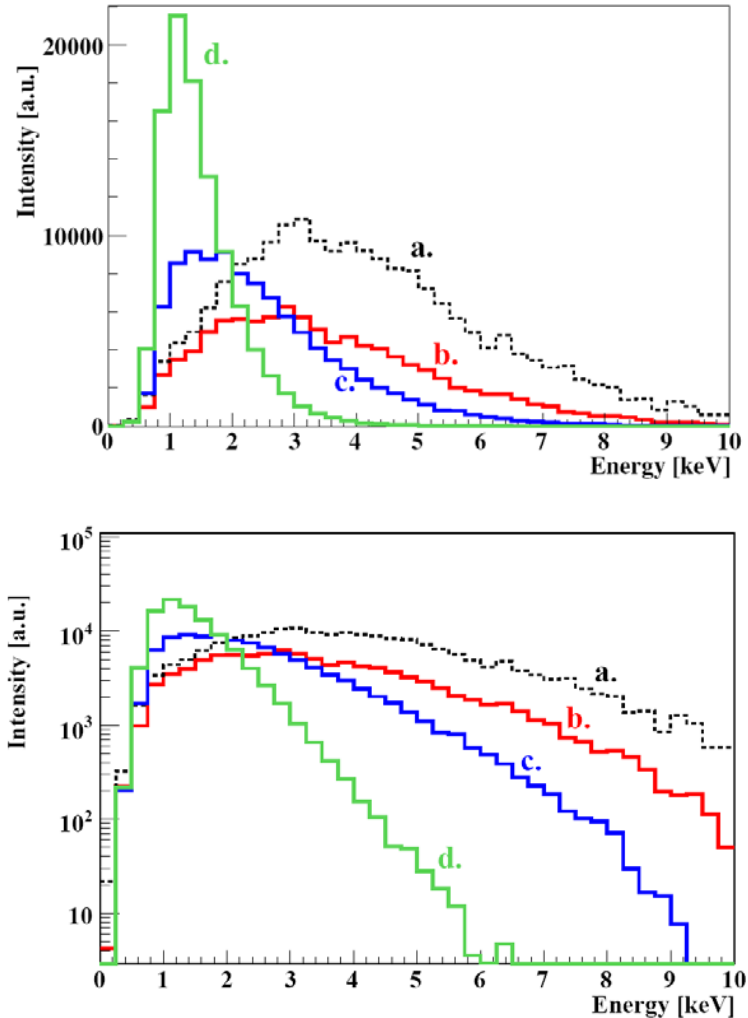

**FIGURE 2** Simulation with the CERN / Geant4 code of the propagation of X-rays from converted solar axions (**a**) at the solar surface magnetic field into space. The converted axions can photoionize the atmosphere above the initial place of birth of the X-ray spectrum emitted radially from magnetically induced axion "decays" according to the inverse Primakoff-effect. The onset of photon energy degradation with increasing atmospheric column density is apparent: (**b**) 4 g/cm$^2$, (**c**) 16 g /cm$^2$, (**d**) 49 g/cm$^2$. The Geant4 code used here follows photons down to 1 keV, i.e. the turnover at ~1 keV is an artifact of the simulation. (Compare these spectra with that given in FIGURE 3.)

## 4. Comparison with observation

It must be noted that there is actually not a measured solar X-ray analog spectrum, say, between ~0.1 and 10 keV, which is apparently of direct interest to check this work. The reasons behind this have to do with the dynamic range, threshold, sensitivity, etc. of the detectors used in space, plus the unpredictably changing solar intensity with time. However, there are few relevant spectra found in ref. [25], which are reconstructed ones for the quiet and flaring Sun (see Figure 3 and 4). Interestingly, both spectra peak towards low energies. The one for the flaring sun (Figure 4) is the mostly relevant one for the reasoning of this work. Figure 4 shows the observed (=reconstructed) solar photon spectrum (**a.**), the expected solar axion spectrum (**b.**) along with the escaping X-rays from axions converted at a depth of ~200 km (16 g/cm$^2$) and ~430 km (64 g/cm$^2$).

The two chosen depths are potential birth places of the assumed axion conversion inside the solar (near surface) magnetic field (inverse Primakoff effect). The estimated plasma energy at these places is ~0.01-0.02 eV, which implies a similar axion rest mass for the resonance axion-photon conversion to take place.

Thus, the Monte Carlo simulation shows that the suggested standard axion picture can explain the observed non-directivity as well as the analog spectrum of the X-ray flaring Sun that peaks at low energies. It cannot predict, however, the X-ray intensity, even though various candidate observations [5-7] follow the striking $B^2$-dependence. That is to say, 3 out of 4 basic axion related properties are encountered in the flaring/magnetic Sun, being thus in favour of the suggested axion scenario.

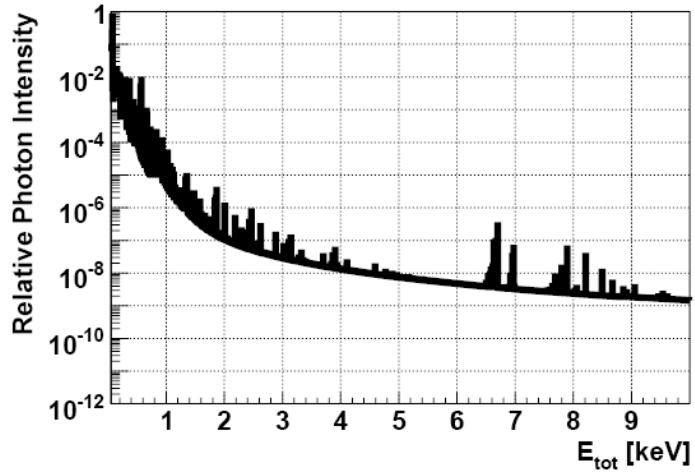

**FIGURE 3** Reconstructed solar photon spectrum from the quiet sun at solar minimum (see figure 9 in ref. [9]). The low energy part (<2 keV) reflects the mysterious solar corona problem. The comparison with the log-scale of FIGURE 2 is suggestive also about the possible origin of this soft X-ray spectrum component from the quiet Sun. With no other solar axion conversion mechanism one could arrive before to such a dominating low energy component. This explains also the observed $B^2$-dependence noticed in [5-7] for X-rays below ~3 keV, which covers the diminishing low energy axion spectrum.

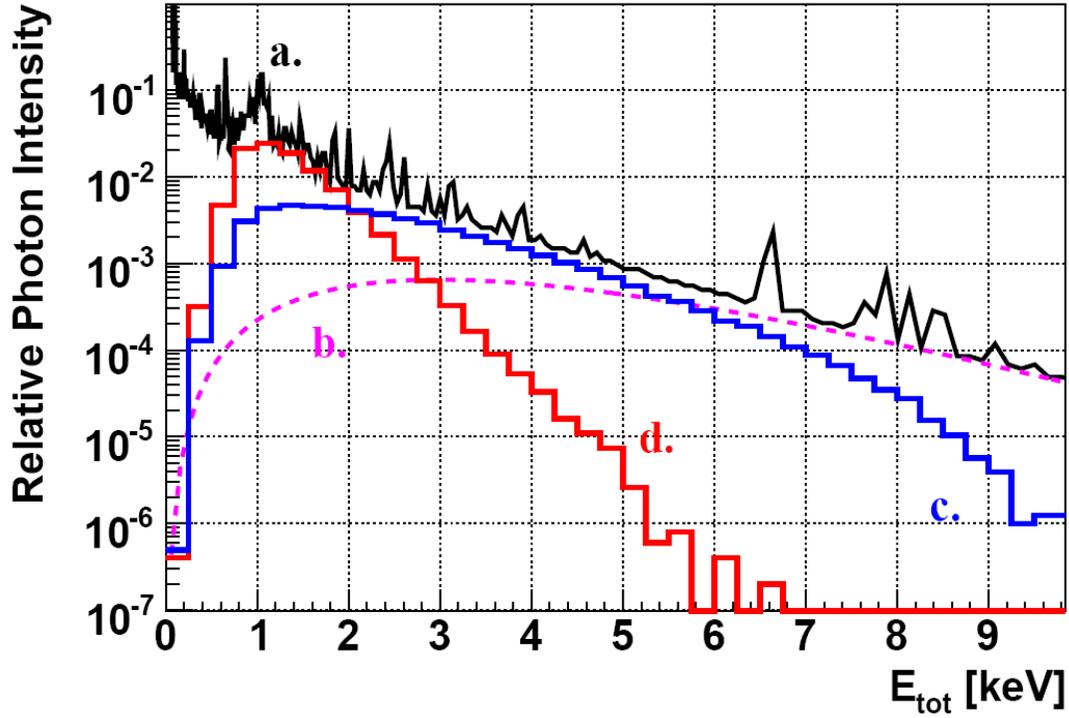

**FIGURE 4**  Flaring Sun: interpretation of a reconstructed solar photon spectrum (**a.**) below 10 keV derived from Figure 6 of ref. [25]. The dashed line (**b.**) is the converted solar axion spectrum assuming zero column density. Two degraded spectra due to multiple Compton scattering are also shown for comparison reasons: (**c.**) 16 g/cm$^2$ and (**d.**) 64 g/cm$^2$. Spectra **b.**, **c.** and **d.** are not to scale. All this fits also the observed B$^2$-dependence in [5-7] for X-rays below ~3 keV, where the initial axion spectrum is negligible compared to the observed one, provided the low energy X-rays originate from converted axions. Note, the used Geant4 code applies above 1 keV. An extrapolation of the obtained Monte Carlo spectra (**c.** and **d.**) towards lower energies is encouraging also for the sub-keV range, where most of the (quiet as well as active) solar irradiance occurs.

## 5. Discussion

While the region above an axion-initiated flaring region is being progressively photoionized, a single Compton scattering of an axion-related hard X-ray photon is sufficient to completely erase its memory of the initial collinearity with the converted axion. This is a possibly important result. Because, combining the isotropic Compton scattering with the degree of modification of the initial analog spectrum (see Figure 1 and 2), it can be used to extract the depth into the photosphere of the initial axion conversion. Furthermore, assuming only the coherent inverse Primakoff effect to be at work, and, the axions are coming from the hot solar core, then the required resonance between axion rest mass and plasma energy maximizes the interaction. This reasoning arrives at axions with a rest mass, within a factor of 2-3, equal to about 0.01 eV/$c^2$, which could materialize somewhere in the ~500km thick photosphere. Within the same reasoning it is natural that X-ray flares and CMEs are observed often together. Under the rather realistic conditions given above, the main results of the suggested axion scenario are:

a) the solar disk centre is no longer exceptional, i.e., it is not the only place for converted axions, streaming out from the core, to be seen by an outside observer; it shows even less X-ray activity.
b) the initial axion energy transferred entirely to the photon can become quite reduced during its propagation inside the thick atmosphere above, before it escapes into space. The spectrum shape can be used to conclude about the conversion place, i.e., axion rest mass.
c) the concluded axion rest mass is ~$10^{-2}$eV/$c^2$, and
d) the coherent inverse Primakoff-effect cannot reproduce the observed large X-ray intensities.

Note that a) and b) were actually used in the past to widely dismiss possible axion involvement in flares. The reasoning of this work favors just the opposite. To put it differently: *if a flare or any other solar activity launched anywhere near its "thick" magnetized surface is due to converted axions or the like, its X-rays can be observed by any observer, and eventually at (much) lower energies, and both fit solar observation.* Remarkably, solar activity takes place preferentially in strong magnetic places, i.e., in sunspots, which appear equally in all longitudes, but are confined within about ±35° in latitude. Therefore, Figure 2 in ref. [17], which was reconstructed from released soft X-ray data taken with the Japanese Yohkoh mission ~11 years ago, might be just one typical overlooked example among many others published before and which were never considered to contain signature(s) of solar axions. Interestingly, the recently made observation with Microflares [18], the distribution of their birth place follows also the magnetic Sun.

It is worth clarifying that the disk centre region [4] could still be occasionally a distinguished one, if the axion conversion process occurs quite high in the "thin" solar atmosphere. So far, such a situation does not fit the bulk of existing solar observations, except the two candidate observations we could find so far in published work [7,16]. A comparison between the behavior of strong X-ray flares and orders of magnitude weaker solar X-ray emission from active regions could provide new and interesting insight.

## 6. Conclusions

Challenging questions like the origin of the soft and hard solar X-ray emission remain elusive within conventional astrophysics. It has been known for many years that the magnetic field plays a crucial role in heating the solar corona, with the exact energy storage and release mechanism(s) being still a nagging unsolved problem for solar physics [15,21]. In this work we show that the solar axion scenario can explain strong solar X-ray activity surpassing the problems associated with the "dogmatic" radial directivity of the emerging photons, and, the observed low energy radiation that dominates solar irradiance spectra. Such signatures have not been previously considered at all as candidates for axions, also for an even simpler reason: axion-related X-ray sources should be rather faint. However, at the moment no conclusion can be drawn about the very nature of the suspected exotica. Even less can we enter into their ultimate interaction strength and/or the main enhancement mechanism, which, in accordance with observation, must be at work preferentially in strong magnetized places near the Sun's surface with a minimum of column density above the actual trigger place of a few $g/cm^2$. However, solar observations are pointing to particles like the celebrated axions or the like, as the first choice for new physics involvement in the (active) Sun, at both large and small scales.

Within this framework, a first Monte Carlo simulation shows that if outstreaming solar axion(-like) particles are converted to X-rays in magnetic fields anywhere near the photosphere or the chromosphere, they can still reach a near Earth X-ray observatory, and this against the widely accepted axion scenario of ref. [4]. The suddenly appearing of X-rays (which are too hard for the ambient temperature) can suffer a quasi continuous energy degradation *and* loss of memory of their initial trajectory due to (multiple) Compton scattering with the atmospheric environment. Surprisingly, this fits many as yet unnoticed wideband observations, i.e., from the "violent" X-ray flares to some much weaker events [5-7], which are all detected from a large part of, or even from the entire magnetized solar disk. Interestingly, this can be explained (apart from their as yet underestimated intensity!) without necessarily inventing new axion physics; the simulated analog spectra fit for the first time, at least qualitatively, the bulk of the low energy radiation from both quiet and flaring Sun in particular. Then there might be no need to invent a less hot solar axion source outside the hot core; this alternative scheme is still possible, but not our favorite following the reasoning of this work. Within the suggested scenario, we conclude that their rest mass must be about $(1-2)\times 10^{-2} eV/c^2$, if the coherent inverse Primakoff effect is the main conversion mechanism. Therefore, the reevaluation / reconsideration of high statistics solar data in the ~1-10 000 eV range is of potential interest.

Finally, we notice that the reality of axion related solar X-ray emission might have a directivity memory of the initial converted axion somewhere between zero and 100%. Therefore, a longitudinally inhomogeneous solar X-ray emission or flaring distribution (ref. [24] might be of relevance in this respect] could be another kind of imprint of the inner Sun and the solar axion atmospheric fine tuning as long as an alternative conventional explanation fails to apply. Therefore, such observations are of potential interest.

The underestimated X-ray intensities by the conventional axion scenario might point at an interaction component, which we have not included so far in our conventional solar axion scenario. For example, the possible axion coupling to the magnetic field gradient is one possibility, and not necessarily the only or even the real one at work. This we consider at the moment as a generic example. In fact, if magnetic field gradients play a dominant role for the axion-to-photon conversion at the solar magnetic surface, whatever the actual mechanism, then, there is not necessarily a contradiction with the obtained best limits for axion-to-photon coupling strength derived by axion helioscopes [1]. Note that all axion experiments use only magnetic dipole fields. Pending theoretical calculations, this reasoning is certainly suggestive for magnetic quadrupoles as the future axion-photon converters, since they might better mimic solar geometry.

As a last example, we mention the massive solar axions of the Kaluza-Klein type [9]. Their behavior is completely different from the conventional ones: they have a mean velocity of ~0.6c, unlike the formers, which are expected to be super-relativistic. In terms of rest mass, the difference to QCD-axions is even more pronounced.

Interestingly, the reported quiet sun X-ray emission [26] in the 3-6 keV range (>3σ), which is not considered as an upper limit, it is argued that it could be explained as the high energy tail of an X-ray emitting hot plasma. Alternatively, following this work, the suggested spectral degradation for the flaring Sun could be at work even stronger in the quiet Sun. For this to happen, the assumed axion-photon conversion should take place deeper in the magnetized solar surface, which implies lower energy escaping photons due to more Compton scatterings. Note that such a hard X-ray emission can also be due to the spontaneous decay of captured massive axions [9], which is a steady emission, while the magnetically induced axion conversion [this work] should be transient coming mainly from the solar disk depending on the spatiotemporal properties of the plasma and the magnetic field.

**Acknowledgements**

The real support we had from Biljana Lakic in estimating the coherence interaction under solar conditions is very gratefully acknowledged. We thank Tullio Basaglia from the CERN library for providing promptly most of the publications used throughout this work. The work presented here is part of the diploma thesis of M.T. K.Z. thanks CERN for long years of hospitality and support of all kind. We also thank Susan Leech O'Neale for reading the manuscript. The support we have received from the Greek funding agency GSRT is gratefully acknowledged. This research was partially supported by the ILIAS (Integrated Large Infrastructures for Astroparticle Science) project funded by the EU under contract EU-RII3-CT-2004-506222.